\documentclass[aps,prb,twocolumn,groupedaddress]{revtex4}

\usepackage{graphicx}
\usepackage{dcolumn}

\begin{document}


\title{Absence of hole pairing in a simple $t$-$J$ model on the
  Shastry-Sutherland lattice} 

\author{P. W. Leung}
\email{P.W.Leung@ust.hk}
\author{Y. F. Cheng}
\affiliation{Physics Dept., Hong Kong University of Science and Technology,
Clear Water Bay, Hong Kong}

\date{\today}

\begin{abstract}
The Shastry-Sutherland model is a two-dimensional frustrated spin
model whose ground state is a spin gap state.
We study this model doped with one and two holes on
a 32-site 
lattice 
using exact diagonalization. When $t'>0$, we find that the diagonal
dimer order that exists at half-filling are retained at these moderate
doping levels. No other order is found to be favored on doping. The
holes are strongly repulsive unless the hopping terms are
unrealistically small. Therefore, the existence of a spin gap at
half-filling does not guarantee hole-pairing in the present case.
\end{abstract}


\maketitle

\begin{figure}
\resizebox{8cm}{!}{\includegraphics{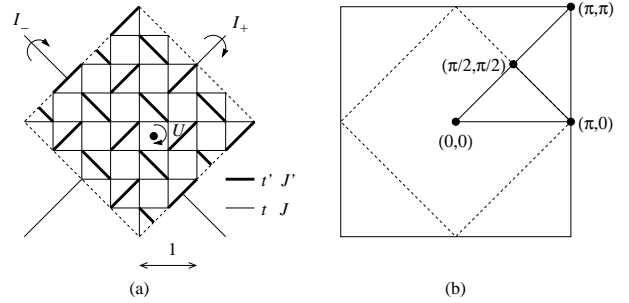}}
\caption{\label{ssl} (a) A 32-site Shastry-Sutherland lattice with
  periodic boundary conditions. Also shown are the two types
  of spin interactions and hole hopping terms. $I_{\pm}$ are
  reflections about the two axis as shown, and $U$ is a four-fold
  rotation about the point indicated by $\bullet$.   (b) The
  corresponding distinct reciprocal space
  vectors.}
\end{figure}

The Shastry-Sutherland (SS) model is an exceptional example of a
two-dimensional
frustrated spin system that has an exact solution.\cite{ss81}
Remarkably, it is also an excellent theoretical model
for the spin gap material SrCu$_2$(BO$_3$)$_2$.\cite{mu03}
The SS model is a Heisenberg model with exchange
interaction $J$ on a two-dimensional square
lattice frustrated by diagonal
bonds $J'$ as shown in
Fig.~\ref{ssl}. When $J/J'$ is smaller than a critical value of about
0.677,[\onlinecite{kk00}]  the
ground state is a direct product of orthogonal singlet dimers residing
on the $J'$ bonds. This is a spin gap state because the
lowest spin excitation involves turning a singlet dimer into a
triplet.  
Different experiments have suggested a range of $J/J'$ for
SrCu$_2$(BO$_3$)$_2$, out of which 0.635 seems to be the optimal one.
A great
deal of work has been devoted to studying the excited states of this
model as an effort to explain various experimental results.
It is found that the almost localized nature of spin excitations
leads to superstructures that give rise to plateaux in the
magnetization curve.
Besides its importance as a theoretical model for
SrCu$_2$(BO$_3$)$_2$, the SS model possesses a quantum phase
transition. One believes that there is at least one intermediate state
between the diagonal dimer state at small $J$ and the N\'eel 
state at large $J$. While the nature of the intermediate state is still
controversial, it seems like the plaquette resonating-valence-bond (RVB)
state and the helical state are among the most likely candidates.
Although discussions on the quantum phase transition are
theoretical, they are not totally irrelevant to
SrCu$_2$(BO$_3$)$_2$. Since its $J/J'$ is not too far from the
critical value, it may be possible to shift it even closer by applying
pressure, substitution, etc. to SrCu$_2$(BO$_3$)$_2$.

The relation between disordered spin liquids with a gap in the spin
excitation and superconductivity has aroused a lot of
interest. It has been suggested that
doping a spin gap system {\it may} lead to hole-pairing and
superconductivity. It is therefore interesting and important to
identify spin gap materials that are Mott insulators with no long-range
spin order. SrCu$_2$(BO$_3$)$_2$ is one such compound.
Although this compound has not been doped, there is no
shortage of suggestions that on hopping the SS model may exhibit
superconductivity. Different types of superconducting states have
been suggested at different doping levels.\cite{sk02,kkaa03,ck03}
One can draw an analogy to a doped two-leg spin ladder.\cite{dr96}
At half-filling the ground state is a spin gap state with singlets forming on
the rungs. When two holes are introduced, they tend to stay on the
same rung in order to avoid breaking more singlets. This is a
beautiful example of a spin gap leading to a mechanism for hole
pairing. The diagonal dimer state of the SS model is also a
spin gap state. But it differs from the two-leg spin ladder in
that it is a two-dimensional model. It is thus
interesting to see whether the same spin gap induced pairing mechanism
is in effect in the SS model. Furthermore, doping may introduce
more interesting physics to the spin background. When $J/J'$
is close to but below the critical value, on doping
the system may select other ordered (or disordered) states that
can better support hole motion.\cite{cms01} Since $J/J'$ for
SrCu$_2$(BO$_3$)$_2$ is close to the critical value, theoretical
discussions on the quantum phase transition may become even more
relevant if it can be doped. 
Here we
report our results on the SS model with up to two holes on a
32-site lattice using numerical diagonalization.

We start from the $t$-$J$ model Hamiltonian,
\begin{equation}
\label{tJ}
{\cal H}=-\sum_{\langle ij\rangle\sigma} t_{ij}(\tilde{c}^\dagger_{i\sigma}
\tilde{c}_{j\sigma}+\tilde{c}^\dagger_{j\sigma}\tilde{c}_{i\sigma})
+\sum_{\langle ij\rangle} J_{ij}\left({\bf S}_i\cdot{\bf S}_j
-\frac{1}{4}n_in_j\right),
\end{equation}
where $t_{ij}=t$ and $J_{ij}=J$ when $\langle ij\rangle$ is a nearest
neighbor pair, and $t_{ij}=t'$ and $J_{ij}=J'$ when $\langle
ij\rangle$ is a diagonal pair as shown in Fig.~\ref{ssl}(a), and zero
otherwise.
Following Ref.~\onlinecite{sk02}, we define $\alpha\equiv t'/t$.
Using the large-$U$ expansion of the Hubbard model as a reference,
namely $J=4t^2/U$, we take $J'/J=\alpha^2$.
Consequently there are only two parameters in our model: $\alpha$
and $J/t$.
Since we want to study hole motion in a dimerized spin
background and to mimic the material SrCu$_2$(BO$_3$)$_2$, we fix
$\alpha=1.25$, i.e., $J/J'=0.64$. The ground state
at half-filling is the SS dimerized state with energy $-(J+J')N/2$
where $N$ is the number of sites.
This leaves us with only one
adjustable parameter $J/t$ when holes are added.
Without guidance from experimental data,
we choose $J/t=0.3$ which is the appropriate value for
high $T_c$ materials. It is also the same choice as in
Ref.~\onlinecite{sk02}.
Using exact diagonalization we solve this model on a 32-site SS
lattice with periodic boundary conditions as shown in
Fig.~\ref{ssl}(a). The size of a unit cell is defined to be 1, i.e.,
nearest neighbor sites are at distance 1/2 apart.
Since there are four sites in a unit cell, an $N$-site SS lattice has
$N/4$ translation operations. The allowed momenta are shown in
Fig.~\ref{ssl}(b). 
Spin reflection symmetry $I_\sigma$ can be used to reduce the size of the
Hilbert space by a factor of 2  when the number of spins is even.
In addition, the SS lattice possesses $C_{4v}$ (or $4mm$) point group
symmetry.\cite{knetter00}  
Therefore eigenstates
can have $s$, $p$, or $d$ symmetries, just like the case on a
square lattice.
In our implementation, we choose to classify eigenstates according to
their symmetry properties with respect to the two reflections
$I_{\pm}$ and the four-fold rotation $U$ as shown in
Fig.~\ref{ssl}(a).[\onlinecite{mm03}] 
Table~\ref{energies} shows the ground state properties of the model
with zero, one and two holes.
 \begin{table}
 \caption{\label{energies} Ground state energies and symmetries of the
   model with
   different number of holes $N_h$ at $\alpha=1.25$ and
   $J/t=0.3$. $N_B$ is the number of basis in that particular symmetry
   subspace.}
 \begin{ruledtabular}
 \begin{tabular}{cdcccccr}
$N_h$ & \multicolumn{1}{c}{$E/t$} & {\bf k} & $I_+$ & $I_-$ & $U$ &
$I_\sigma$ &\multicolumn{1}{c}{$N_B$}  \\
\colrule
0 & -12.3 & $(0,0)$ & $+$ & $+$ & $+$& $+$& 4,708,641 \\
1 & -15.780\,034 & $(0,0)$& $+$ & $+$& $-$ & & 150,297,603 \\
2 & -19.291\,015 & $(0,0)$& $+$ & $+$& $+$ & $-$&  601,144,932\\
 \end{tabular}
 \end{ruledtabular}
 \end{table}

\begin{figure}
\resizebox{7cm}{!}{\includegraphics{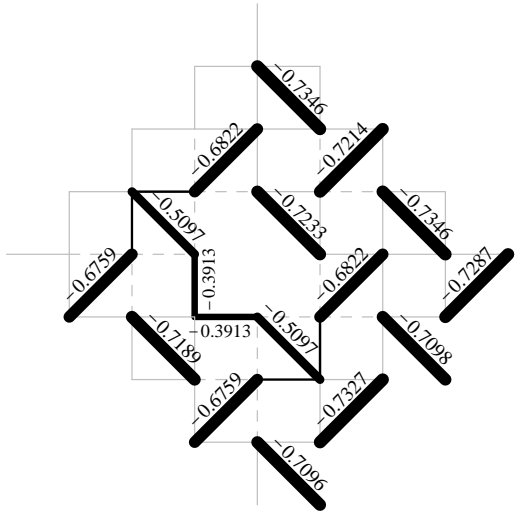}}
\caption{\label{1hspin} Spin structure of the one-hole ground
  state. The hole is at the open lattice point at the
    center of the lattice. Thickness of the lines are proportional to
    the magnitudes of $\langle{\bf S}_i\cdot{\bf 
    S}_j\rangle$, whose values are also shown for strong singlet
    bonds. Solid and broken lines 
    represent negative and positive numbers respectively. Shaded lines
    correspond to correlations too small to be shown in this
    scale. Their thickness do not represent their magnitudes, but are
    draw to show their signs only.}
\end{figure}

The one hole ground state has $d$-like symmetry as shown in
Table~\ref{energies}.
Fig.~\ref{1hspin} shows the spin structure of this ground state. It is
obvious that the introduction of a mobile hole does not alter the
dimerized spin background much. Most of the diagonal bonds remain as
strong singlets with $\langle{\bf S}_i\cdot{\bf S}_j\rangle$ not much
weaker than $-3/4$. 
The most noticeable feature is an almost free five-spin chain formed
next to the hole. When a spin is removed from one dimer, the resulting
free spin tends to 
minimize energy by forming a five-spin chain. In this way the system
avoids the high energy cost of having a free spin and at the same time
minimizes disruption to the spin background.
Note that
$\langle{\bf S}_i\cdot{\bf S}_j\rangle$ along this ``chain''
are not too far from the corresponding values in an isolated five-spin
chain with the same $J/J'$, which are 
$-0.6588$ and $-0.2855$ respectively.
Also note that it can easily
lower its energy through hopping --- the apex  of the chain
can hop to a diagonal site without disrupting the spin
background. The average kinetic energy per bond on the diagonal and
nearest neighbor bonds are $-0.06062\,t'$ and $-0.04597\,t$
respectively, showing that diagonal hopping is more active than
nearest neighbor hopping. Furthermore, this excitation is quite localized.
The
bandwidth is only $0.1611\,t$, much smaller than the case on a square
lattice which is about $0.6\,t$.\cite{lg95}

\begin{figure}
\resizebox{8cm}{!}{\includegraphics{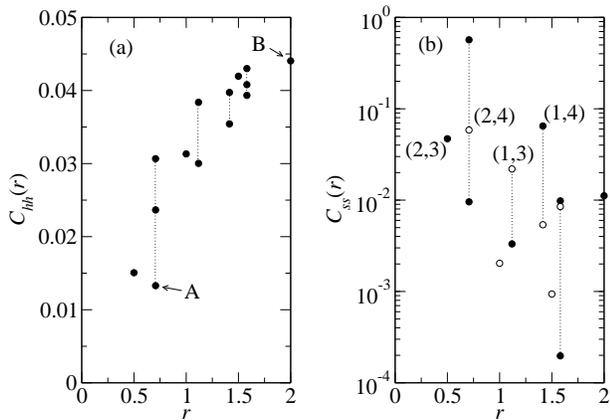}}
\caption{\label{chh} (a) Hole correlation $C_{hh}(r)$ and (b) spin correlation
  $C_{ss}(r)$  of the two-hole model. Dotted lines join inequivalent
  points with the same 
  $r$. In (b), open and filled circles represent positive and negative
  correlations respectively.}
\end{figure}

\begin{figure}
\resizebox{7cm}{!}{\includegraphics{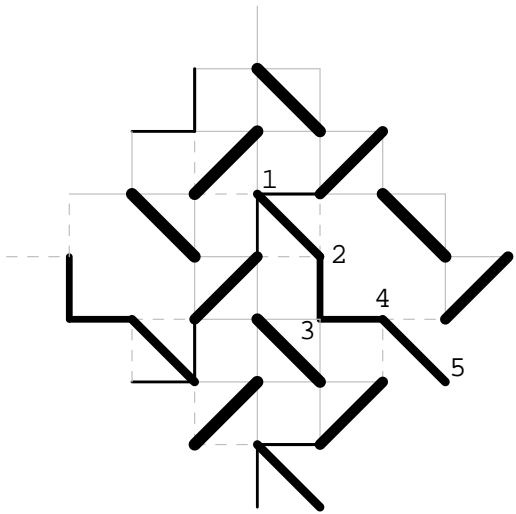}}
\caption{\label{2hspin} Same as Fig.~\ref{1hspin} except for the
  two-hole model projected to the subspace where the holes are at the
  two open lattice points.}
\end{figure}

The two-hole ground state
wavefunction is $s$-like as shown in Table~\ref{energies}. 
The two-hole binding energy
$E_b\equiv E_{2h}-2E_{1h}+E_0$ is  $-0.0309\,t$ at
$J/t=0.3$. Although $E_b$ is negative, its small value makes it
doubtful whether it really represents hole-binding. Fig.~\ref{chh}(a)
shows the hole
correlation function
$C_{hh}(r)\equiv \langle(1-n_r)(1-n_0)\rangle$, which is the
probability of occurrence of that particular hole configuration.
It 
clearly shows that the holes prefer to stay away from each other,
i.e., they are repulsive. 
The configuration corresponding to hole-pairing due to spin gap
effect, where the
holes reside on the same diagonal bond, is indicated by ``A'' in
Fig.~\ref{chh}(a).
Its hole correlation is the smallest. 
The most favorable hole configuration is when they are at maximum distance
apart as indicated by ``B'' in Fig.~\ref{chh}(a).
Fig.~\ref{2hspin} shows its spin structure. A striking
feature is that next to each hole there is an almost free five-spin
chain which we have seen in the one-hole ground state,
Fig.~\ref{1hspin}. 
The spin correlation $C_{ss}(r)\equiv\langle{\bf S}_r\cdot{\bf
  S}_0\rangle$ shown in Fig.~\ref{chh}(b) is also consistent with the
existence of such spin chain. In the diagonal dimer state at
half-filling, $C_{ss}(r)$ is zero everywhere except when the holes are in
the same diagonal 
bond.   $C_{ss}(r)$ at this point
remains the largest in Fig.~\ref{chh}(b),
showing that dimer order
persists in the two-hole ground state. In addition there are four $C_{ss}(r)$
at  $r=1/2, 1/\sqrt{2}, \sqrt{5}/2$, and $\sqrt{2}$
that have appreciable values. They are due to correlations between spins in 
the five-spin chains. 
For illustration purpose we label these four points in
Fig.~\ref{chh}(b)  by  $(a,b)$ where $a$ and $b$ are sites on a five
spin chain such as those marked in Fig.~\ref{2hspin}.
For example, the positive correlation $(2,4)$ in Fig.~\ref{chh}(b) is
due to spins 2 and 4 in Fig.~\ref{2hspin}.
Once we establish the existence of five-spin chains,
 we can easily understand
why holes 
repel. If the holes are well separated such that a
five-spin chain can be formed next to each hole, then the local spin energy
is minimized by getting rid of the otherwise free spins, and kinetic energy is
favored by the hopping motion of the apex of the spin chains.

\begin{figure}
\resizebox{8cm}{!}{\includegraphics{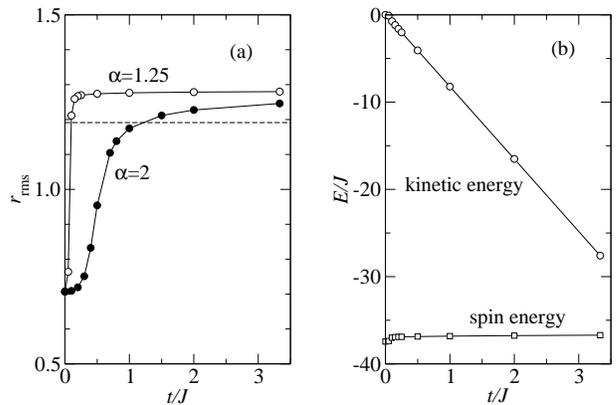}}
\caption{\label{rms} (a) Root-mean-square separation of holes in
  the two-hole ground state at $\alpha=1.25$ and 2. The broken line
  indicates the value for a 
  pair of uncorrelated holes on this 32-site lattice. (b) Kinetic and
  spin energies at $\alpha=1.25$.}
\end{figure}

At $t=0$ the two-hole ground state is the exact dimerized state just like at
half-filling but with one dimer removed. Therefore hole pairing must exist in
this limit. Let us look at how it is destroyed as $t$
increases. 
In Fig.~\ref{rms}(a) we plot the root-mean-square separation
of holes $r_{\text{rms}}\equiv\sqrt{\langle r^2\rangle}$ at different
$t/J$. The horizontal broken line shows the value for two uncorrelated holes on
this 32-site lattice, which is 1.1914. 
Let us consider the result at $\alpha=1.25$ first.
At
$t/J=0$, the holes 
are at the same diagonal bond and $r_{\text{rms}}=1/\sqrt{2}$. But as
$t/J$ increases, $r_{\text{rms}}$ increases very
fast and at $t/J\sim 0.1$ it already exceeds the uncorrelated value.
When $t/J\agt 0.2$, $r_{\text{rms}}$ becomes saturated and the holes
become strongly repulsive. In Fig.~\ref{rms}(b) we plot the kinetic
and spin energies which are the ground state expectation of the first
and second terms of ${\cal H}$ in Eq.(\ref{tJ}) respectively.
Except at very small $t/J$ where $r_{\text{rms}}$ is
still increasing rapidly, the kinetic energy decreases
linearly with $t/J$ while the spin energy barely increases.
This shows that once the system enters the hole-repelling regime,
kinetic energy of the holes dominates and their motion do not cause
much disruption to the spin background. This is in total agreement
with our picture discussed above.

For the purpose of comparison we also show the corresponding
result at $\alpha=2$ in Fig.~\ref{rms}(a). Note that this corresponds
to $J/J'=0.25$, which is far from the realistic parameter for
SrCu$_2$(BO$_3$)$_2$. With a much larger spin gap --- $3.7172J$ compared to
$0.6015J$ at $\alpha=1.25$, the resulting hole-binding effect persists to
larger $t/J$, and $r_{\text{rms}}$ reaches the uncorrelated value when
$t/J$ is slightly larger than 1. However, this represents only a
qualitative change. The kinetic effect eventually wins and the
holes are unbound when $t/J>1$.

\begin{figure}
\resizebox{7cm}{!}{\includegraphics{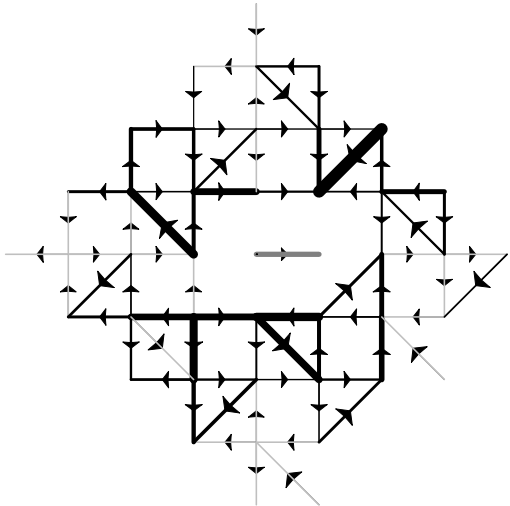}}
\caption{\label{currentcorr} The current correlation $\langle
  j_{kl}j_{mn}\rangle$ in the two-hole model. 
  The reference bond is denoted by a
  shaded line with an arrow near the center of the diagram. 
  Thickness of solid lines are proportional to the magnitudes
  of the current correlations. Thin shaded lines represent
  correlations that are too small to be shown in this scale.}
\end{figure}

So far our results show that  diagonal dimer order is retained on
doping. Next we try to look for ``hidden'' orders that may be
associated with  charge motion.
Our previous study\cite{pwl00} has shown that the $d$-wave state of
the two-hole $t$-$J$ model on a square lattice
has a
staggered pattern in the  current correlation $\langle
j_{kl}j_{mn}\rangle$, where the current on a bond linking sites $k$ and
$l$ is defined by
$j_{kl}=it_{kl}(\tilde{c}^\dagger_{k\sigma}\tilde{c}_{l\sigma} -
\tilde{c}^\dagger_{l\sigma}\tilde{c}_{k\sigma})$.
This staggered current pattern is closely related to the
staggered-flux phase based on the SU(2) theory.\cite{ilw00}
On the SS lattice, a similar staggered-flux phase
has been suggested by a mean-field study\cite{ck03}. However, we know
that on a square lattice the staggered current pattern exists only  in
the $d$-wave state where the holes are attractive. But in the present
case the ground state has $s$, not $d$ symmetry, and the holes are
repulsive. Hence we do not expect to see a staggered current pattern in this
ground state. Fig.~\ref{currentcorr} clearly shows  no
staggered pattern in the current correlation. Therefore we do not find
evidence for a staggered-flux phase in this model.

Not only does dimerized order persists on doping, we believe that
doping in fact stabilize it near the quantum phase
transition point. To illustrate this we
repeat our calculations at $\alpha=1.2$, i.e., $J/J'=0.694$ which is
beyond the quantum phase transition point.  The
ground state at half-filling on a 32-site SS lattice shows plaquette
order in the dimer correlation.\cite{lws02} 
However, on doping with one and two holes the system dramatically
change back to diagonal dimer order.
The spin structures and correlations $C_{hh}(r)$ and $C_{ss}(r)$ are
qualitatively the same as in Figs.~\ref{1hspin},
\ref{2hspin} and \ref{chh} respectively. 
The one- and two-hole ground state energies are
$-15.214\,070\,t$ and $-18.736\,734\,t$, giving a binding
energy  $E_b=-0.0206\,t$, compared to $-0.0309\,t$ at $\alpha=1.25$. 
$r_{\text{rms}}$ is 1.2822, compared to
1.2800 at $\alpha=1.25$. Since all results are so similar, our
previous conclusions apply to the 
present case with $\alpha=1.2$, i.e.,
the holes are repulsive and the spin structure shows
strong diagonal dimer order. 
Since the plaquette RVB state exists at most in a very small range of
$J/J'$, the fact that on doping it change back to diagonal dimer order
means that doping destroys plaquette RVB order and stabilize diagonal
dimer order.

To summarize, we find that doping the dimerized ground state of the SS
model does not favor any other ordered or disordered state. In fact
doping favors diagonal dimer order. 
When a mobile hole is
introduced to the dimerized state, the ``free spin'' that would have been
formed minimize energy by forming a five-spin chain with two of its
nearest dimers. This excitation is localized and has a small
dispersion compared to the $t$-$J$ model on a square lattice.
When more than one hole is present, their respective spin chains tend
to avoid each other. This give rise to short-range repulsion between
the holes. Although there may exist long-range attraction between the
holes mediated through the chains, 
Fig.~\ref{rms}(b) shows that the spin energy is not
sensitive to changing spin configurations. Therefore we consider
long-range attraction unlikely. There is another competing short-range pairing
mechanism --- 
the spin gap naturally leads to pairing of holes on the same diagonal
bond when $t=0$. But we find that this pairing mechanism is in
effect only when $t/J$ is smaller than about 0.2 at $\alpha=1.25$. 
Note that this value
corresponds to $U/t\simeq 4t/J=0.8$, which is far from the strongly
interacting regime of the Hubbard model. As a result, the idea of
treating the dimerized ground state of SrCu$_2$(BO$_3$)$_2$ as a Mott
insulating state of an underlying Hubbard model will
fail. Consequently we consider $t/J<0.2$ too small to be
realistic. 
The implication of our result is that the existence of spin
gap in a disordered spin liquid state does not guarantee charge binding.

Finally, we remark that our results do not completely rule out
hole pairing in the SS model because we have studied the case
$t'>0$ only.
Just like in the case of a square lattice, the sign of $t$ is
irrelevant. 
On a square lattice, $t'$ breaks electron-hole symmetry and 
N\'eel order of 
the spin background is better preserved with $t'>0$.
This property is consistent with the 
phase diagram of electron-doped high $T_c$
superconductors.\cite{tohyama} Similarly, we have shown that on a SS lattice
diagonal dimer order of the spin background persists on doping when
$t'>0$. But without experimental results we cannot decide whether
$t'>0$ corresponds to electron-doping in the present
case. Nevertheless, it is meaningful to study the case $t'<0$ where
diagonal hopping is discouraged.
These works are in progress and will be
presented elsewhere.


This work was supported by the Hong Kong RGC grant number
HKUST6075/02P. Most calculations were performed on a 64-CPU AMD
Opteron cluster computer in the Physics Department of HKUST.

\end{document}